\documentclass[10pt,twocolumn]{IEEEtran}
\renewcommand{\baselinestretch}{1}
\usepackage{amsmath}
\usepackage{cite}
\usepackage{graphicx}
\usepackage{tabularx}
\usepackage{anyfontsize}
\usepackage{lettrine}
\usepackage{forloop}
\usepackage{algorithm}
\usepackage{algorithmic}
\usepackage{array}

\DeclareMathOperator{\atanh}{atanh{}}
\DeclareMathOperator{\sign}{sign{}}
\allowdisplaybreaks
\begin{document}

\title{Design of Capacity Approaching Ensembles of LDPC Codes for Correlated Sources using EXIT Charts}
\author{Mohamad~Khas, Hamid~Saeedi,~\IEEEmembership{Member,~IEEE,} and Reza~Asvadi,~\IEEEmembership{Member,~IEEE} 
	\thanks{M. Khas and H. Saeedi are with the Department of Electrical
		Engineering, Tarbiat Modares University, Tehran, Iran (e-mail:
		hsaeedi@ieee.org).}
	\thanks{R. Asvadi is with the Faculty of Electrical Engineering, Shahid Beheshti University (SBU),
		Tehran, Iran. (e-mail: r\_asvadi@sbu.ac.ir).} }

\maketitle

\begin{abstract}
This paper is concerned with the design of capacity approaching ensembles of Low-Densiy Parity-Check (LDPC) codes for correlated sources. We consider correlated binary sources where the data is encoded independently at each source through a systematic LDPC encoder and sent over two independent channels. At the receiver, a iterative joint decoder consisting of two component LDPC decoders is considered where the encoded bits at the output of each component decoder are used at the other decoder as the a priori information.  We first provide asymptotic performance analysis using the concept of extrinsic information transfer (EXIT) charts. Compared to the conventional EXIT charts devised to analyze LDPC codes for point to point communication, the proposed EXIT charts have been completely modified to able to accommodate the systematic nature of the codes as well as the iterative behavior between the two component decoders. Then the developed modified EXIT charts are deployed to design ensembles for different levels of correlation. Our results show that as the average degree of the designed ensembles grow, the thresholds corresponding to the designed ensembles approach the capacity. In particular, for ensembles with average degree of around 9, the gap to capacity is reduced to about 0.2dB. Finite block length performance evaluation is also provided for the designed ensembles to verify the asymptotic results.

\end{abstract}
\vspace{-.6 cm}

\section{Introduction}\label{Introduction}

Source and channel encoding/decoding of correlated sources has been the subject of several studies \cite{slepian1973noiseless,
veaux2000channel,garcia2001joint,adrat2001iterative,poulliat2005analysis}. Perhaps the most immediate example of correlated sources is in sensor networks in which each sensor measures the data, encode it to bits and transmits it to a central node for decoding \cite{akyildiz2002survey,chong2003sensor,barros2006network}. The correlation of the encoded bits comes from the fact that in many cases, several sensors measure the same phenomenon. The closer the sensors are, the larger the degree of  correlation will become in most cases.
 On the other hand, due to energy limitation in sensors which is enforced to increase the sensor lifetime, it is essential that the data is transmitted with the lowest possible energy while maintaining the required bit error rate. Consequently, channel coding is usually deployed at each sensor prior to transmission. At the central node, the channel decoder should be applied to each block of data received from each sensor. Now if the received bit streams are correlated, it is natural to consider joint channel decoding to take advantage of such a correlation.

Low-density Parity-check (LDPC) codes \cite{gallager1962low} have been widely suggested to be deployed in sensor networks due to their remarkable performance and reasonable decoding complexity \cite{zhu2009distributed,shahid2011distributed,shahid2013distributed,xiong2004distributed,monteiro2016multi,deligiannis2015distributed,sartipi2008distributed}. For point to point communications, sequences of capacity approaching ensembles over memoryless Gaussian channels have been proposed in \cite{richardson2001design, saeedi2011successive} where for a given code rate, the threshold of the ensemble is numerically shown to approach the Shannon capacity as the average check node degree increases. For the binary erasure channels, capacity achieving sequences of ensembles have been designed and their thresholds have been analytically shown to achieve the capacity \cite{oswald2002capacity,saeedi2010new}.

For point to point LDPC codes, different tools and techniques have been deployed for ensemble design. The most well-known analysis tool is density evolution (DE) \cite{richardson2001capacity}. To reduce the design complexity, an important alternative tool known as Extrinsic Information Transfer (EXIT) chart was proposed in \cite{ten2004design}, \cite{ashikhmin2004extrinsic} based on the assumption that the exchanged messages between variable nodes (VN) and check nodes (CN) of the corresponding Tanner graph can be approximated by consistent\footnote{For a consistent Gaussian random, the variance of the distribution is twice its mean.} Gaussian random variables.

%

\begin{figure}[!t]
\centering
\includegraphics[scale=0.29]{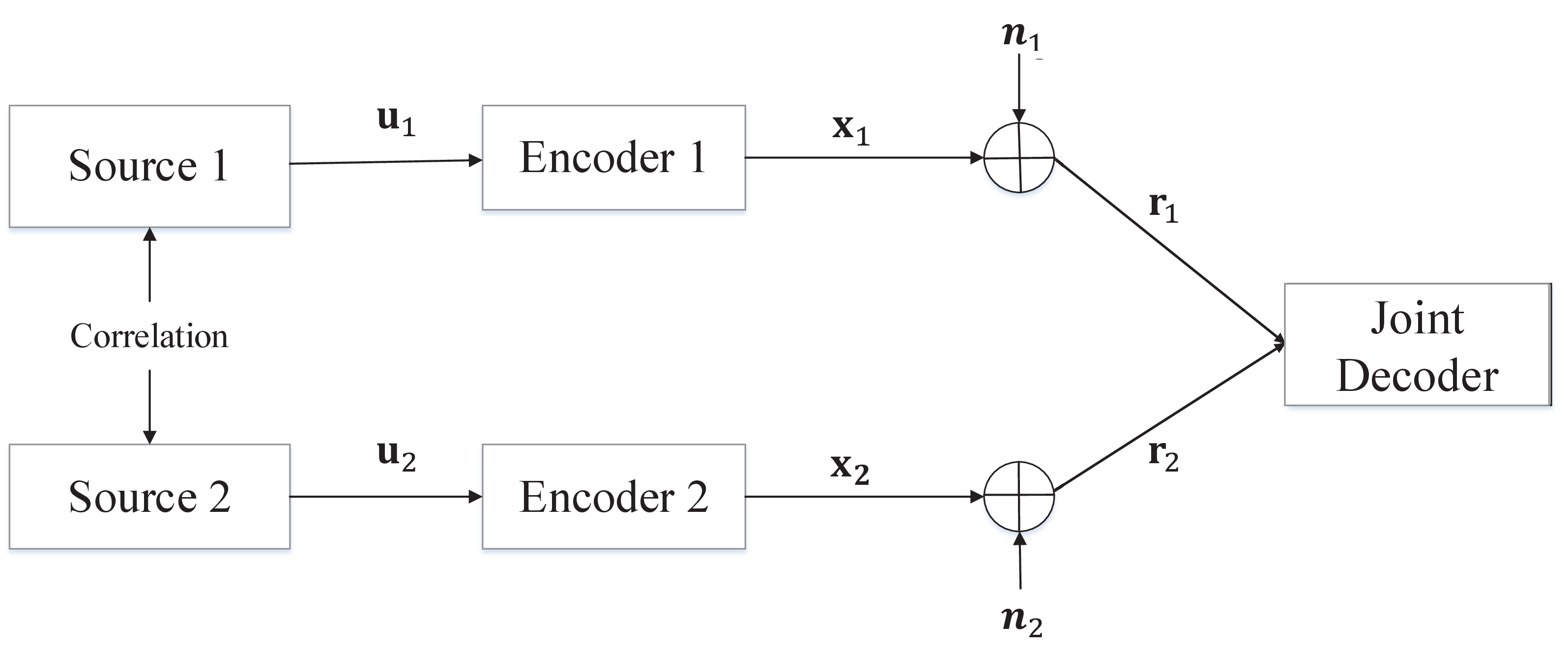}
\caption{Block diagram of the system model}\label{F-system model}
\end{figure}

In this paper, we consider the problem of joint channel decoding of LDPC-encoded correlated binary sources. We consider a simplified model where two sources generate
correlated binary bit streams. Then the streams are fed to an LDPC encoder blockwise and sent through two independent additive white Gaussian noise (AWGN) channels as shown in Fig. \ref{F-system model}. Then the streams of data are received blockwise by the central node and are fed-back to the joint LDPC decoder proposed in \cite{daneshgaran2006ldpc} to obtain the original bit streams. In this decoder, two types of iterations, namely, inner and outer iterations are deployed such that at each outer iteration, the output of one decoder is used as the \emph{a priori} information of the other decoder while the inner iterations are performed similar to a conventional LDPC message passing decoder. 

Our aim in this paper is to design capacity approaching ensembles of LDPC codes in which we show that the decoding threshold of the joint decoding of the designed ensembles tends to the capacity limit obtained in \cite{garcia2001joint} as the average check node degree of the designed ensembles grow. We in fact obtain tables of degree distributions (similar to those of \cite{richardson2001design} proposed for the point to point scenario) for different levels of correlation in the source bits for the first time. The claimed capacity approaching thresholds are then verified by finite block length simulations.

To design the ensembles, we use the concept of EXIT charts. There are, however, important obstacles to apply the original EXIT chart scheme to our case that are addressed in this paper. First, as there are two decoders in place, the EXIT curve corresponding to the variable node of each decoder should be generated taking into account the \emph{a priori} information from the other decoder. Second, as we consider systematic codes, the corresponding Tanner graph in our case has a two edge-type structure \cite{ning2008design}, one corresponding to the message bits and one corresponding to the  parity bits. Therefore, two types of variable node EXIT curve have to be considered and the corresponding degree distributions have different structure than the conventional non-systematic LDPC codes. We show that with reasonable average check node degree, we can get as close as 0.2dB of the Shannon limit for different amount of correlations.


The organization of the paper is as follows. In Section \ref{Source and Correlation Model}, basic concepts and notations related to the source and correlation model and the Shannon limit are given. In Section \ref{system model}, we describe the algorithm for iterative joint channel decoding of correlated sources and propose the two edge-type structure for this model. In Section \ref{EXIT-CHART ANALYSIS}, we propose the modified EXIT charts for joint iterative LDPC decoding of correlated sources and analyse the performance of regular and irregular LDPC codes. The code design procedure is presented in Section \ref{optimization}. The simulation results and numerical examples are summarized in Section \ref{finite result}. Finally, Section \ref{Conclusion} draws the conclusion.

\section{Preliminaries}\label{Source and Correlation Model}
\subsection{Source and Correlation Model }

Consider the two binary memoryless sources $(U_{1},U_{2})$ that generate binary sequences segmented in blocks of length $K$ and denoted by $\boldsymbol{{u}}_1=\{u_{1,1},u_{1,2},\dots,u_{1,K}\}$ and $\boldsymbol{{u}}_2=\{u_{2,1},u_{2,2},\dots,u_{2,K}\}$.
The bits within each sequence are assumed to be i.i.d with equal probability of being zero and one. \cite{yedla2013code,daneshgaran2006ldpc,sartipi2008distributed}.

Let $\boldsymbol{{z}}=\boldsymbol{{u}}_1\oplus\boldsymbol{{u}}_2$ be the component-wise addition modulo-2 of the two sources output. The vector $\boldsymbol{z}=(z_1,\dots,z_k)$ implies correlation between the two sources and it is called correlation vector. We define the empirical correlation between these two sources as $p =\gamma /K$ where $\gamma$ is the number of zeros in $\boldsymbol{{z}}$. Obviously, sequences with the empirical correlation values of $p$ and $1-p$ have the same \textit{entropy} value \cite{daneshgaran2006ldpc}. This correlation can be generated by simply passing tone of the sequences through a binary systematic channel (BSC) with transition probability $1-p$ to generate the sequence for the other source. 

A joint channel decoding for the correlated sources is considered, where the sources are independently encoded by identical LDPC encoders, i.e., encoders have no communication. The encoders map a $k$ bit vector corresponding to $\boldsymbol{{u}}_1$ ($\boldsymbol{{u}}_2$) to $n_1$ ($n_2$) bit vector of $\boldsymbol{{x}}_1$ ($\boldsymbol{{x}}_2$). The code rate of each encoder is then equal to $RR_{c_1}=K/n_1$.
 and $R_{c_2}=k/n_2$. In what follows, we assume that the code rates are the same (\textit{symmetric system}), i.e, $R_{c_1}=R_{c_2}=R_c$ , or equivalently, $n_1=n_2=n$
\cite{daneshgaran2006ldpc,yedla2013code}. Each source is encoded according to a systematic $(n,K)$ LDPC code. Hence, the generated codeword $\boldsymbol{{c}}$ is the augmentation  of information bit  $\boldsymbol{{u}}$ and parity bit $\boldsymbol{{p}}$ vectors. The binary phase-shift keying (BPSK) modulation is used before sending a codeword $\boldsymbol{c}$ over the AWGN channel. 

\subsection{Theoretical limit}\label{Theoretical limit}
To be able to evaluate the performance of the joint-decoding, the \textit{Shannon-SW} limit is considered \cite{garcia2001joint}.
In our simulations, we employ the energy per generated source, denoted by $E_{so}$, which is related to the energy per information bit, denoted by $E_b$, and the energy per transmitted symbol, denoted by $E_s$, as follows \cite{garcia2005near, garcia2007turbo}:
\begin{equation}
2E_{so}=H(U_1,U_2)E_b=(1/R_{c_1}+1/R_{c_2})E_s,
\end{equation}
where $H(U_1,U_2)$ represents the joint entropy between the two correlated sources and $E_s$ equals $1$ in BPSK modulation. Since $U_1$ and $U_2$ are  uniformly distributed binary sources, the binary entropy of each source is equal 1, i.e, $H(U_1)=H(U_2)=1$. Then, $H(U_1|U_2)=H(U_2|U_1)=h_2(p)$, where $h_2(p)$ is the entropy of the empirical correlation $p$ and $H(U_1,U_2)=H(U_1)+h_2(p)$. In the considered \textit{symmetric system}, reliable transmission over a channel pair is possible as long as $E_{so}/N_0$ satisfies the \textit{Shannon-SW} condition \cite{garcia2001joint}
\begin{equation}\label{shannon-SW-limit}
\frac{E_{so}}{N_0}>\frac{1}{R_c}(2^{H(u_1,u_2)R_c}-1),
\end{equation}
where $N_0$ is the noise power spectral density.

\section{SYSTEM MODEL}\label{system model}

\subsection{Two edge-Type LDPC Codes }\label{Source and Correlation Model}
In this section, a \emph{two edge-type} LDPC code and its associated graph are presented. Consider the parity check matrix $\boldsymbol{H}_{m\times n}$ of an LDPC code represented by a \emph{Tanner} graph \cite{tanner1981recursive}, denoted by $\mathcal{G}=(\mathcal{V},\mathcal{C},\mathcal{E})$, where $\mathcal{V}$ and $\mathcal{C}$ denote a set of $n$ VN and $m$ CN, respectively, and $\mathcal{E}$ is a set of edges of the graph. According to $\mathcal{G}$, we have an edge $e=\{v_i,c_j\}\in \mathcal{E}$, where a VN $v_i \in \mathcal{V}$ is connected to a CN $c_j \in \mathcal{C}$ in $\mathcal{G}$, if and only if, $h_{i,j}=1$ in $\boldsymbol{H}=[h_{i,j}]$.

Conventional LDPC codes are described asymptotically by coefficient pairs $(\lambda,\rho)$ or degree distribution polynomials of the VN and CN as follows:
\begin{equation}
\lambda (x)=\sum_{i=2}^{D_v} \lambda_i x^{i-1}, ‎\quad‎ \rho (x)=\sum_{j=2}^{D_c} \rho_j x^{j-1},
\end{equation}
where $D_v$ and $D_c$ are the maximum VN and CN degrees. The coefficient $\lambda_i$ (resp. $\rho_i$) is the fraction of edges that are connected to VNs (resp. CNs) of degree-$i$.

LDPC codes can be used in a systematic form, and hence its codeword comprises two disjoint \emph{source} and \emph{parity} parts. Accordingly, the VNs are divided into two sets: \emph{source nodes} and \emph{parity nodes}. Then, the edges connecting to source or parity nodes are called \emph{source edges}, denoted by $\mathcal{E}^s$, and \emph{parity edges}, denoted by $\mathcal{E}^p$, respectively. In this paper, we use a family of LDPC codes whose CNs are connected to \emph{source nodes} via at least one edge. The LDPC codes with such a structure are called \emph{fully-source-involved} LDPC (FSI-LDPC) codes.

\begin{figure}[!t]
	\centering
	\includegraphics[scale=0.45]{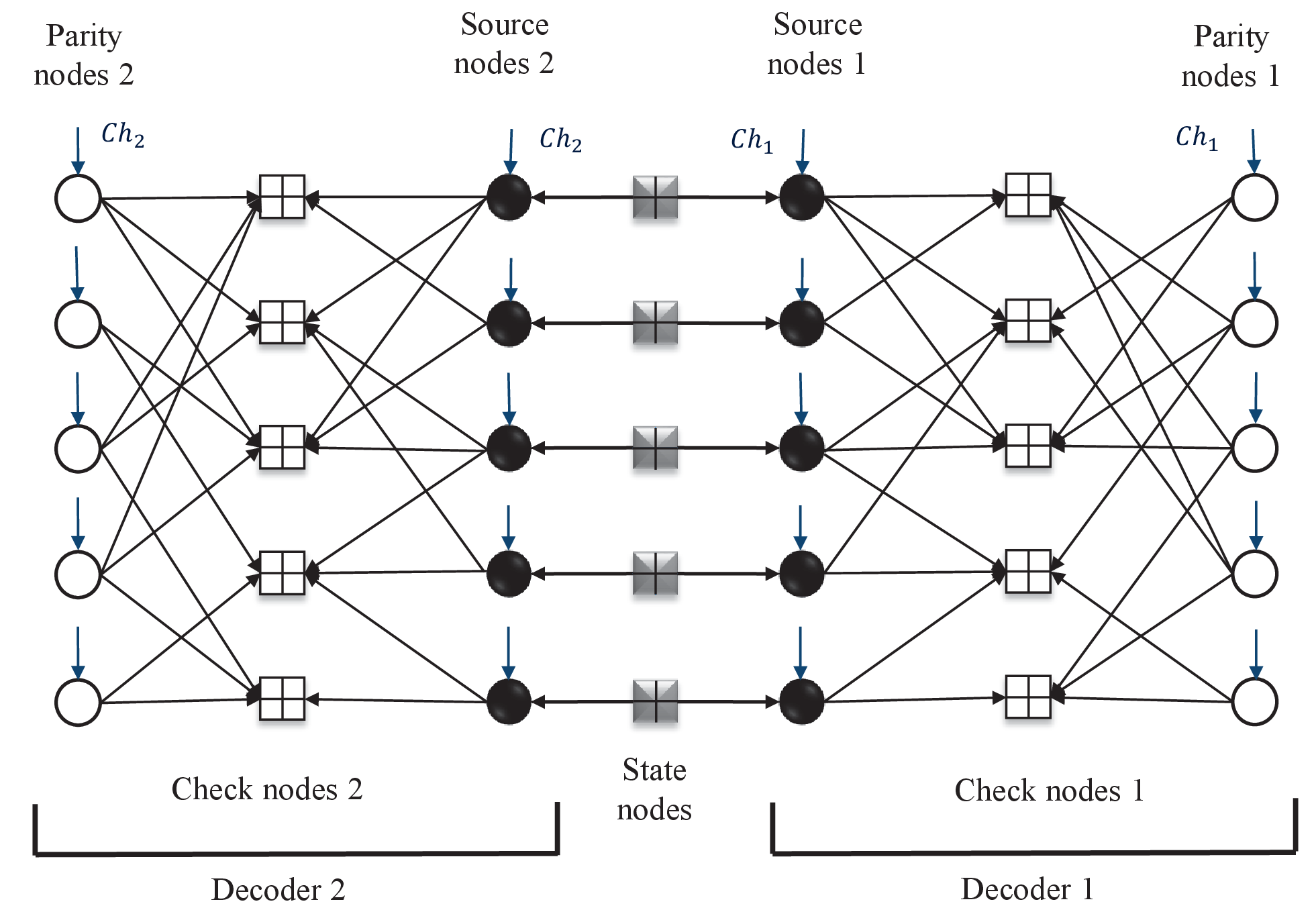}
	\caption{ A schematic of the \emph{two edge-type} graph of an iterative joint channel decoder. Dark circle and white circle nodes depict source and parity nodes, respectively. Gray and white square nodes also depict state and CNs, respectively. }\label{F-multi-edge}
\end{figure}

Since a joint decoder employs two compound LDPC decoders, the associated Tanner graph of the joint decoder consists of three types of nodes. They are called \emph{source nodes} denoted by $\mathcal{V}^s=\{\textit{v}^s_1,\textit{v}^s_2,\dots,\textit{v}^s_{n-m}\}$, \emph{parity nodes} denoted by $\mathcal{V}^p=\{\textit{v}^p_1,\textit{v}^p_2,\dots,\textit{v}^p_{m}\}$, and CNs denoted by $\mathcal{C}$ for each of LDPC decoders. Moreover, there are \emph{state nodes} denoted by $\mathcal{S}$ which connect the two iterative decoders to exchange extrinsic information. A schematic of the associated \emph{Tanner} graph of the joint decoder is presented in Figure \ref{F-multi-edge}. In this figure, information of \emph{source nodes} of each decoder are passed to the \emph{source nodes} of the other decoder via \emph{state nodes}. Furthermore, we use $\mathcal{G}=(\mathcal{V}^s,\mathcal{V}^p,\mathcal{C},\mathcal{S},\mathcal{E}^s, \mathcal{E}^p)$ to denote a \textit{two edge-type} graph of a joint LDPC decoder.

We follow the notations defined in \cite{ning2008design} in this paper. Let $n^s_i$ and $n^p_i$ denote the number of \emph{source} and \emph{parity} nodes of degree-$i$, respectively. The total number of \emph{source} and \emph{parity} nodes are also given by $n^s=\sum_{i=2}^{D_v}n^s_i$ and $n^p=\sum_{i=2}^{D_v}n^p_i$, respectively. Let $n_i=n^s_i+n^p_i$ denote the number of VNs of degree-$i$. Hence, the number of VNs $n$ equals $\sum_{i=2}^{D_v}n_i$. Let $m_{j,k}$ be the number of CNs of degree-$j$, $k$ edges of which are connected to the \emph{source} nodes and $(j-k)$ edges of which are connected to the \emph{parity} nodes. Thus, the number of CNs of degree-$j$, denoted by $m_j$, is equal to $\sum_{k=1}^{j-1}m_{j,k}$. Similarly, the total number of CNs $m$ is determined by $\sum_{j=2}^{D_c} m_j$, and the total number of edges on a \textit{two edge-type} graph is determined by
\begin{equation}
E=E^s+E^p=\dfrac{n}{\sum_{i=2}^{D_v}\lambda_i/i}=\dfrac{m}{\sum_{j=2}^{D_c}\rho_j/j}.
\end{equation}

In addition to $(\lambda,\rho)$, we need to introduce additional coefficient pairs $(\alpha,\beta)$ to asymptotically describe a \textit{two edge-type} graph where $\alpha_i=n^s_i/n_i$ is in fact the fraction \emph{source} nodes of degree-$i$ out of all degree-$i$ VNs. Similarly, $\beta_{j,k}=m_{j,k}/m_j$, where $\sum_{k=1}^{j-1}\beta_{j,k}=1$. Now, the \emph{source} and \emph{parity} variable degree distribution polynomials are, respectively, defined as follows:

\begin{equation}
\lambda^s (x)=\sum_{i=2}^{D_v} \lambda_i^s x^{i-1}, ‎\quad‎ \lambda^p (y)=\sum_{i=2}^{D_v} \lambda_i^p y^{i-1},
\end{equation}
where
\begin{equation}
\lambda_i^s=\dfrac{n_i^s\cdot i}{E^s}=\dfrac{n_i^s}{n_i} \  \dfrac{n_i\cdot i}{E} \ \dfrac{E}{E^s} = \frac{\alpha_i \lambda_i}{\sum_{j=2}^{D_v} \alpha_j \lambda_j},
\end{equation}

\begin{equation}
\lambda_i^p=\dfrac{n_i^p\cdot i}{E^p}=\dfrac{n_i-n_i^s}{n_i} \  \dfrac{n_i\cdot i}{E} \ \dfrac{E}{E^p}=\frac{(1-\alpha_i) \lambda_i}{\sum_{j=2}^{D_v} (1 - \alpha_j) \lambda_j}.
\end{equation}
The \emph{source} and \emph{parity} side CN degree distribution polynomials are, respectively, defined as follows:
\begin{equation}
\rho^s (x,y)=\sum_{j=2}^{D_c} \sum_{k=1}^{j-1} \rho_{j,k}^s  x^{k-1}  y^{j-k},
\end{equation}

\begin{equation}
\rho^p (x,y)=\sum_{j=2}^{D_c} \sum_{k=1}^{j-1} \rho_{j,k}^p  x^{k}  y^{j-k-1},
\end{equation}
where

\begin{equation*}
\rho_{j,k}^s = \dfrac{m_{j,k}\cdot k}{E^s} = \dfrac{\rho_j}{j}\cdot \frac{\beta_{j,k} k}{\sum_{i=2}^{D_v} \alpha_i \lambda_i},
\end{equation*}

\begin{equation*}
\rho_{j,k}^p = \dfrac{m_{j,k}\cdot (j-k)}{E^p} = \dfrac{\rho_j}{j}\cdot \frac{\beta_{j,k} (j-k)}{\sum_{i=2}^{D_v} (1-\alpha_i) \lambda_i}.
\end{equation*}

There are two variables denoted by $``x"$ and $``y"$ in $\rho^s (x,y)$ and $\rho^p (x,y)$, which indicate two types of edges incident to each CN. The inner summation of $\rho_{j,k}^s$ and $\rho_{j,k}^p$ over $k$, indicate the fraction of $j$-th degree CNs which are connected to the the \emph{source} and \emph{parity} VNs, respectively.
The code rate must be the same in \textit{two edge-type} and its corresponding \textit{single edge-type} graph. Furthermore, the number of edges emerging from each type of variable and CNs must be the same too. Hence,the following conditions \cite{ning2008design} must be satisfied:
\begin{equation}\label{condition-alpha-beta}
\sum_{i=2}^{D_v} \alpha_i \lambda_i=\sum\limits_{j=2}^{D_c}\dfrac{\rho_j}{j} \sum\limits_{k=1}^{j-1} \beta_{j,k}k,
\end{equation}
\begin{equation}\label{condition-alpha}
\sum_{i=2}^{D_v} \dfrac{\alpha_i \lambda_i}{i}=R \sum_{i=2}^{D_v} \dfrac{\lambda_i}{i}.
\end{equation}

Note that a symmetric \emph{two edge-type} graph corresponding to a joint decoder can be described by $(\lambda, \rho, \alpha, \beta)$.
Moreover, an ensemble corresponding to $(\lambda, \rho, \alpha, \beta)$ can be obtained from a \textit{single edge-type} ensemble $(\lambda, \rho)$ by partitioning the VNs and their associated edges connected to the CNs according to $(\alpha,\beta)$ distribution. 

\subsection{Iterative Joint LDPC Decoding }\label{joint decoder}
We assume that the data block  corresponding to sources $U_1$ and $U_2$, i.e, $\boldsymbol{u}_1$ and $\boldsymbol{u}_2$ are encoded  through the systematic LDPC encoders respectively into $\boldsymbol{{x}}_1$ and $\boldsymbol{{x}}_2$. The encoded bits are transmitted over two independent AWGN channels. At the receiver, we receive vectors $\boldsymbol{{r}}_1=\boldsymbol{{x}}_1+\boldsymbol{n}_1$ and $\boldsymbol{{r}}_2=\boldsymbol{{x}}_2+\boldsymbol{n}_2$,
where $\boldsymbol{n}_i \sim \mathcal{N}(0, \sigma_n^2)$, for $i=1,2$, 

Let $\boldsymbol{\hat{{u}}_i}$, $i=1,2$, denote decoded bits of $i$-th decoder. The joint receiver employs the empirical estimate of the correlation parameter to benefit from the intra-sources correlation. The estimate of correlation vector, denoted by $\boldsymbol{\hat{{z}}}$, is calculated by $\boldsymbol{\hat{{z}}}=\boldsymbol{\hat{{u}}}_1\oplus \boldsymbol{\hat{{u}}}_2$.


\begin{figure}[!t]
\centering
\includegraphics[scale=0.73]{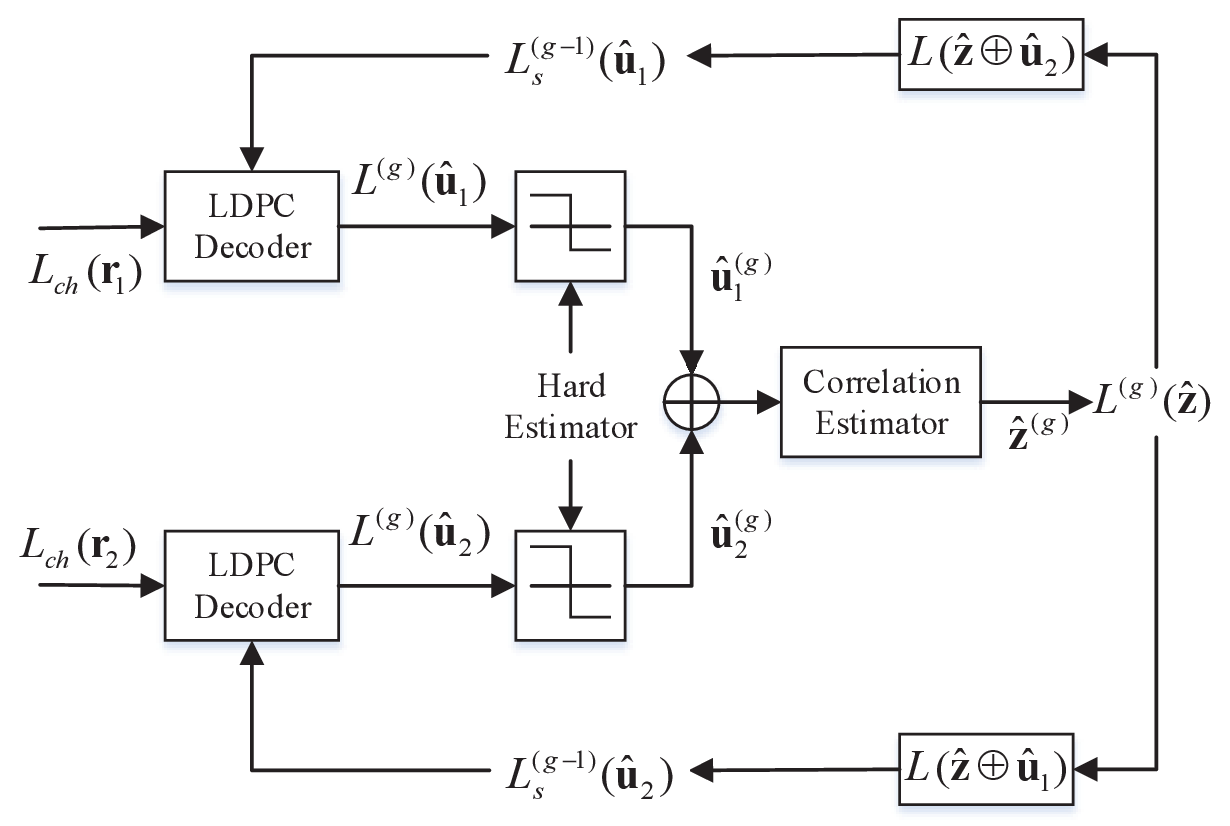}
\caption{Block diagram of the iterative joint channel decoder of correlated sources}\label{F-joint decoder}
\end{figure}
The joint decoder is composed of two parallel LDPC decoders working based on the \textit{sum-product} (SP) algorithm \cite{mackay1999good} where it also accepts an estimate of the transmitted bits from the other decoder as a priori information. The structure of iterative joint decoder is shown in Fig. \ref{F-joint decoder}.
There are tow types of iterations, which are called \emph{global} and \emph{local} iterations indicated by superscripts $g$ and $l$, respectively. The estimate of correlation is updated during each global iteration. Then, this update estimation is passed on to the both decoders for being used as a side information. Next, each decoder performs the SP algorithm with a specified maximum number of local iterations. Note that log-likelihood ratio (LLR) of the side information are only added to the systematic bit nodes of each decoder, while these LLRs are first set to zero for the both decoders.
Consider the $g$-th global and $l$-th local iterations of the joint decoder. In each local iteration, the encoded bits coming from the channel are transformed to \emph{a posteriori} LLRs, denoted by $L_{ch}(\boldsymbol{r})$, and fed into the VNs as follows:

\begin{equation} \label{llr-channel}
L_{ch}(r_{i,v})=\log(\dfrac{\Pr(x_{i,v}=1|r_{i,v})}{\Pr(x_{i,j}=0|r_{i,v})})=\dfrac{2}{\sigma^2}r_{i,v},
\end{equation}
where $i \in \{1,2\}$ indicates the $i$-th channel, and $v \in \{1,2,\dots,n\}$ is the number of VN. For simplicity, we drop the channels index in the sequel unless an ambiguity arises. Let $L_{v,c}^{(l)}$ and $L_{c,v}^{(l)}$ denote the LLR of messages emanating from a VN $v$ to a CN $c$ and, vice versa, from the CN to the VN at local iteration $l$, respectively. Since the side information are only added to the information bit nodes, LLR update equations from these nodes to the CNs are given by:

\begin{equation}\label{LLR-information nodes}
L_{v,c}^{(l)}=L_{ch}(r_v)+\sum\limits_{c'\neq c}L_{c',v}^{(l-1)} + L_{s}^{(g-1)}(\hat{u}),
\end{equation}
where $v\in \mathcal{V}^s$, and $L_{s}^{(g)}(\hat{u})$ denotes LLR of the side information associated to the estimated correlation at the global iteration $g$. It is worth noting that the summation is only applied on all connected CNs to the VN $v$, except the CN $c$. The messages forwarded from parity VNs to the CNs are calculated in a same way as for the standard SP decoder \cite{mackay1999good}

\begin{equation}\label{LLR-parity nodes}
L_{v,c}^{(l)}=L_{ch}(r_v)+\sum\limits_{c'\neq c}L_{c',v}^{(l-1)},
\end{equation}
where $v\in \mathcal{V}^p$. Furthermore, the messages $L_{c,v}^{(l)}$ are updated in each local iteration in the same equation as represented in the standard SP decoder \cite{mackay1999good}.

In the $g$-th global iteration, the correlation vector $\hat{\boldsymbol{{z}}}$ is estimated by $\hat{\boldsymbol{z}}^{(g)}=\hat{\boldsymbol{u}}_1^{(g)}\oplus \hat{\boldsymbol{u}}_2^{(g)}$, where $\hat{\boldsymbol{u}}_1^{(g)}$ and $\hat{\boldsymbol{u}}_2^{(g)}$ are, respectively, the hard estimates of the source bits $\boldsymbol{u_1}$ and $\boldsymbol{u}_2$ in the terminated local iteration $l_t$ in each associated decoder. The LLR $L^{(g)}(\hat{z})$ at each global iteration is obtained using the proposed technique in \cite{daneshgaran2006ldpc}, as follows:

\begin{equation}
L^{(g)}(\hat{z}_v)=(1-2\hat{z}_v^{(g)})\log_2 (\dfrac{k-W_H}{W_H}),
\end{equation}
where $k$ and $W_H$ are, respectively, the source block size and Hamming weight of the correlation vector $\hat{\boldsymbol{z}}^{(g)}=(\hat{z}_1^{(g)},\dots,\hat{z}_k^{(g)})$.

Finally, the LLR of side information of the source bits that are input to the first and the second decoders, at the next global iteration, are calculated by \cite{hagenauer1996iterative}:
\begin{multline}
L_{s}^{(g)}(\hat{u}_{1,v})=\sign{\left(L^{(g)}(\hat{z}_v)\right)}\cdot \sign{\left(L^{(g)}(\hat{u}_{1,v})\right)} \\
\cdot 2 \atanh{\left(\tanh{(|L^{(g)}(\hat{z}_v)|/2)} \tanh{(|L^{(g)}(\hat{u}_{2,v})|/2)}\right)},
\end{multline}
where $v\in \mathcal{V}^s$, and similarly
\begin{multline}
L_{s}^{(g)}(\hat{u}_{2,v})=\sign{\left(L^{(g)}(\hat{z}_v)\right)}\cdot \sign{\left(L^{(g)}(\hat{u}_{2,v})\right)} \\
\cdot 2 \atanh{\left(\tanh{(|L^{(g)}(\hat{z}_v)|/2)} \tanh{(|L^{(g)}(\hat{u}_{1,v})|/2)}\right)},
\end{multline}
where
\begin{align}
& L^{(g)}(\hat{u}_{1,v})=L_{ch}(r_{1,v})+\sum\limits_{c'}L_{c',v}^{(l_t-1)},\\
& L^{(g)}(\hat{u}_{2,v})=L_{ch}(r_{2,v})+\sum\limits_{c'}L_{c',v}^{(l_t-1)},
\end{align}
and $v \in \mathcal{V}^s$. According to the \emph{Turbo principle}, side information, also called \emph{extrinsic} information, added to each SP decoder should not include the same decoder's information. Therefore, side information from $L(\boldsymbol{\hat{u}}_1)$ and $L(\boldsymbol{\hat{u}}_2)$ are removed from the two above equations, but they have to be added when information source bits are estimated at the end of the local iteration to calculate \emph{a posteriori} information.

\section{EXIT CHART ANALYSIS}\label{EXIT-CHART ANALYSIS}
In EXIT chart analysis tool firstly proposed in \cite{ashikhmin2004extrinsic} to design ensembles of conventional LDPC codes, evolution of the Mutual Information (MI) between a given transmitted binary bit at source and several LLR's within the decoder is traced. In a point to point (P2P) LDPC decoder, variable node EXIT curve displays the MI between the transmitted bit and the LLR at the output of the VN ($I_{EV}$) versus the MI between the transmitted bit and the LLR at the input of the VN ($I_{AV}$). Similarly, check node EXIT curve displays the MI between the transmitted bit and the LLR at the output of the CN ($I_{EC}$) versus the MI between the transmitted bit and the LLR at the input of the CN ($I_{AC}$).

To obtain these curves, it is usually assumed that the density at the input of each decoding unit (VN and CN in this case) are consistent Gaussian random variables. To track the iterative exchange of messages between VN and CN, we set $I_{AV}$ at iteration $l$ equal to $I_{EC}$ of iteration $l-1$.  Similarly, we set $I_{AC}$ at iteration $l$ equal to $I_{EV}$ of iteration $l$. Alternatively, one can plot the VN curve and inverse of the CN curve ad following the trajectories. This is referred to as EXIT chart. As far as the EXIT chart analysis of the proposed system model is concerned, we have to deal with a modified chart that can incorporate both inner and outer iterations as well as the fact the considered graph is \textbf{\emph{two edge-type}} in contrast to the conventional case.

\subsection{Mutual Information between the Transmitted Bit and the LLR of the Received Data}\label{EXIT-CHART Definition}
Let $X_1$ be a BPSK modulated transmitted bit in the first source and $X_2$ be the BPSK modulated transmitted bit at the second source on the same time instant. Since the sources are correlated, we have $P(X_1=X_2)=p$. Now let $A_1$ and $A_2$ be the LLRs corresponding to the received information at the input of the first and second decoders, respectively. It has been already established that $A_i$ are consistent Gaussian random variables with variance $\sigma^2_A=4/\sigma^2$. It has been shown that \cite{ashikhmin2004extrinsic} in this case we have $I(X_c,A_d)=J(\sigma^2_A), c=d, c=1, 2$ where
\begin{multline}\label{J-equ}
J(\sigma_A)=\\
1-\frac{1}{\sqrt{2\pi}\sigma_A} \int_{-\infty}^{\infty} {\log_2(1+e^{-l}) \exp(-\frac{(l-\sigma_A^2/2)^2}{2\sigma_A^2})dl}.
\end{multline}
Moreover, for $c \neq d$, we have $I(A_c;X_d) =\tilde J(\sigma_A,p)$ \cite{razi2014convergence} where
\begin{multline}\label{J-tilde}
\tilde J(\sigma_A,p)=1-\\
\frac{1}{\sqrt{2\pi}\sigma_A} \int_{-\infty}^{\infty} \log_2(\frac{1+e^{-l}}{p+\bar pe^{-l}})
\left(pe^{\tiny -\frac{(l-\sigma_A^2/2)^2}{2\sigma_A^2}} + \bar pe^{\tiny -\frac{(l+\sigma_A^2/2)^2}{2\sigma_A^2}} \right ) dl,
\end{multline}
and $\bar{p}=1-p$.

If the correlation parameter is assumed to $ 100\% $, i.e., $p = 1$, Eq. (\ref{J-tilde}) is reduced to the well-known equation (\ref{J-equ}).
 Likely, if the correlation parameter is set to be $ 50\% $, i.e., $ p=0.5 $, then MI of the \emph{extrinsic} messages passed to the other decoder would be equal to zero which is intuitively reasonable.
Eq. (\ref{J-tilde}) will be used in the next subsection to incorporate the correlation between $X_1$ and $X_2$ in the corresponding EXIT chart.
\subsection{Modified EXIT Chart}\label{modified-EXIT}

Consider one of the decoders at the receiver. At outer iterating $g$, the variable node EXIT curve of degree $i$ belonging to source (systematic) bits at inner iteration $l$, $I_{EV}^{s(l)}(i)$, can be obtained  based on the data from channel, extrinsic information from the check node at iteration $l-1$ and, extrinsic information coming from the other decoder at outer iteration $g-1$. The latter term is referred to as the helping information and denoted by $I_{h}^{g}$. $I_{h}^{g}$ is in fact a function of $I_{EV}^{s(l_t)}$ of the other decoder at outer iteration $g-1$ where $l_t$ is the last inner iteration. At first outer iteration, this is set to zero.

Moreover, at outer iterating $g$, the variable node EXIT curve belonging to parity bits at inner iteration $l$, $I_{EV}^{p(l)}(i)$, can be obtained  based on the data from channel and extrinsic information from the check node at iteration $l-1$. Similar statements can be made for the check node exits curves.
Given the above explanations and using (\ref{LLR-information nodes}), we obtain $I_{EV}^{s(l)}(i)$ as
\begin{align}
\begin{split}
I&_{EV}^{s(l)}(i)  = \\
& J \left( \sqrt{\sigma_{ch}^2 + (i-1)[J^{-1}(I_{EC}^{s(l-1)})]^2 + [J^{-1}(I_{h}^{(g-1)})]^2 }    \right),
\end{split}
\end{align}
where $I_{EC}^{s(l-1)} $ denotes the mutual information from the CNs to the \emph{source} nodes at $g$-th global iteration.
So for an irregular variable node the EXIT curve is obtained as:
\begin{equation}\label{MI-INtoCN}
I_{EV}^{s(l)} = \sum_{i=2}^{D_v} { \lambda_i^s I_{EV}^{s(l)}(i) },
\end{equation}
To obtain $I_{h}^{(g-1)}$ we act as follows:
We first obtain $ I_{h}^{(g-1)}(i)$, the $I_{h}^{(g-1)}$ corresponding to degree-$i$ VNs. Note that it is in fact equal to $I(X_c,A_d)$ for $c\neq d$.  Therefore we have
\begin{equation*}
I_{h}^{(g)}(i) = \tilde{J}(\sigma_h,p), \ \text{where} \ \sigma_h = \sqrt{\sigma_{ch}+i[J^{-1}(I_{EC}^{s(l_t)})]^2} \ ,
\end{equation*}
where $I_{EC}^{s(l_t)}$ is the extrinsic information of at the output of the check nodes of the other decoder at last iteration.
We finally obtain:
\begin{equation}\label{MI-H}
I_{h}^{(g)}=\sum_{i=2}^{D_v} { \lambda_i^s  I_{h}^{g}(i) },
\end{equation}

To obtain $I_{EV}^{p(l)}(i)$, using (\ref{LLR-parity nodes}) we write
\begin{align}
\begin{split}
I_{EV}^{p(l)}&(i)  =  J \left( \sqrt{\sigma_{ch}^2 + (i-1)[J^{-1}(I_{EC}^{p(l-1)})]^2 }    \right),
\end{split}
\end{align}
where $I_{EC}^{p(l-1)} $ denote the mutual information from the CNs to the \emph{parity} nodes at $g$-th global iteration.
So for an irregular variable node the EXIT curve is obtained as:
\begin{equation}\label{MI-PNtoCN}
I_{EV}^{p(l)} = \sum_{i=2}^{D_v} { \lambda_i^p I_{EV}^{p(l)}(i) },
\end{equation}
where $ I_{EC}^{s(l)} $, and $ I_{EC}^{p(l)} $ denote MI from the CNs to the \emph{source} and the \emph{parity} nodes at $g$-th global iteration, respectively. Let $ \gamma_s $ and $ \gamma_p $  be defined as ratio of source edges to total edges and parity edges to total edges respectively that can be written as
\begin{equation}
\gamma_s=\frac{E^s}{E} =\sum_{i=2}^{D_v} \lambda_i \alpha_i, \qquad  \gamma_p= \frac{E^p}{E}=\sum_{i=2}^{D_v} \lambda_i (1-\alpha_i).
\end{equation}
Using (\ref{MI-INtoCN}) and (\ref{MI-PNtoCN}), the VN EXIT curve at outer iteration $g$ is finally obtained as
\begin{equation}\label{MI-VNtoCN}
I_{EV} =  \gamma_s I_{EV}^{s} + \gamma_p I_{EV}^{p}.
\end{equation}

To obtain EXIT curve of the CNs, 
we first obtain $ I_{EC}^{s(l)} $ and $ I_{EC}^{p(l)} $ as follows. Consider degree $j$ CNs which are connected through $k$ edges the \emph{source} nodes and $(j-k)$ edges to the \emph{parity} nodes. The corresponding MI to \emph{source} and \emph{parity} nodes are, respectively, given by:

\begin{align}
\begin{split}
& I_{EC}^{s(l)}(j,k)=1 - \\
& J \left(\sqrt{ (k-1) [J^{-1}(1-I_{EV}^{s(l)})]^2 + (j-k) [J^{-1}(1-I_{EV}^{r(l)})]^2 } \right),
\end{split}
\end{align}
and
\begin{align}
\begin{split}
& I_{EC}^{p(l)}(j,k)=1 - \\
& J \left( \sqrt{ k [J^{-1}(1-I_{EV}^{s(l)})]^2 + (j-k-1) [J^{-1}(1-I_{EV}^{r(l)})]^2 } \right),
\end{split}
\end{align}
Consequently we have:
\begin{equation}\label{MI-CNtoSN}
I_{EC}^{s(l)}=\sum_{j=2}^{D_c} \sum_{k=1}^{j-1}{ \rho_{j,k}^s \  I_{EC}^{s(l)}(j,k) },
\end{equation}
and
\begin{equation}\label{MI-CNtoPN}
I_{EC}^{p(l)}=\sum_{j=2}^{D_c} \sum_{k=1}^{j-1}{ \rho_{j,k}^p \  I_{EC}^{p(l)}(j,k) }.
\end{equation}
Therefore, the overall EXIT curve of the CNs at is obtained as follows:
\begin{equation}\label{MI-CNtoVN}
I_{EC}=\gamma_s  I_{EC}^{s} + \gamma_p  I_{EC}^{p}.
\end{equation}
From (\ref{MI-CNtoSN}) and (\ref{MI-CNtoPN}), we observe that similar to the P2P case, the CN curve is only concerned with the degree of CNs and dose not change with SNR of the channel. Moreover, evaluation of (\ref{MI-CNtoVN}) shows that the CN curve independent from $g$ which also makes sense intuitively.

%
%

\subsection{Modified EXIT chart example}
To get a better insight out of the above ugly equations, the \emph{modified EXIT chart} corresponding to a joint decoder with a $(3,6) $-regular LDPC code is depicted in Fig. \ref{F-EXIT36} for the correlated sources when the correlation parameter $p$ is set to $0.95$ where the maximum inner iterations are set to be 50. Inner and outer iterations have also been plotted. As can be seen, in the absence of the helping information at the first outer iteration, each decoder performs inner iterations but it is stuck in the intersection of the check node curve and the blue variable node curve. Then, at the second outer iteration, with the help of extrinsic formation of the other decoder, the inner iterations continue and the MI continues to grow until it is stuck again at the intersection of check node curve and the red variable node curve. In the 3rd outer iteration, the MI jumps to continue its path toward 1 on the yellow variable node curve.
%
\begin{figure}[!t]
\centering
\includegraphics[scale=0.6]{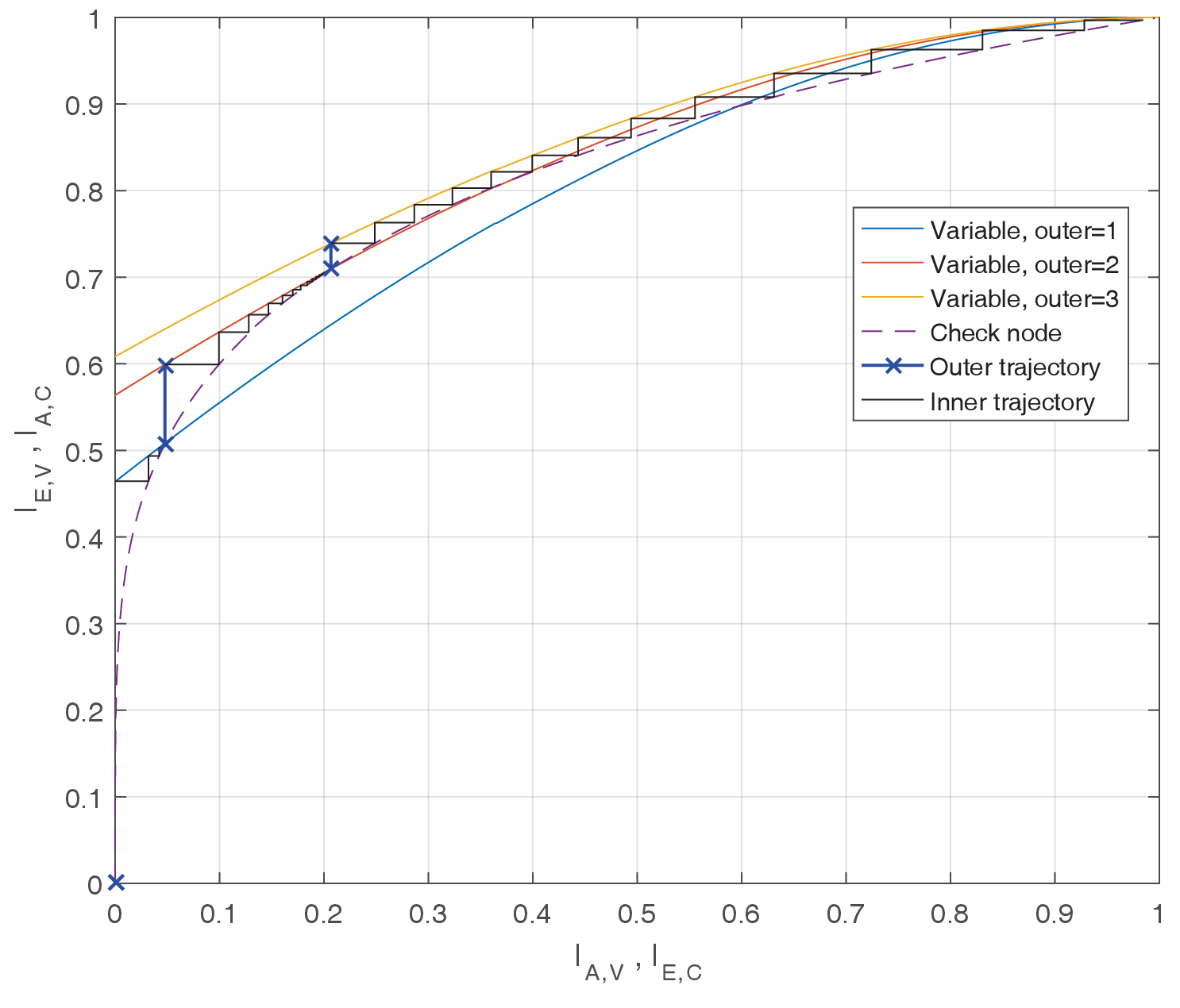}
\caption{ A modified EXIT chart example for a $(d_v,d_c) = (3,6)$ regular LDPC code ensemble with the correlation parameter $p=0.95$ and $\alpha_3 = 0.5$ on a BI-AWGN channel with $E_{so}/N_0=-0.3 $ dB.}\label{F-EXIT36}
\end{figure}

\section{CODE DESIGN FOR THE JOINT DECODER}\label{optimization}
In this section, our aim is designing VN degree distribution $ \lambda(x) $ for an ensemble of LDPC codes with a given CN degree distribution $ \rho(x) $ so as to minimize the gap between code rate $ R $ and the \emph{Shannon-SW} limit corresponding to the channel parameter $\sigma^2$. In general, the design of irregular LDPC codes using EXIT chart analysis is based on a curve-fitting method including EXIT curves of the VNs and the CNs, see, e.g., \cite{ashikhmin2004extrinsic}, \cite{ten2004EXIT}. To make the design procedure simpler, it is often assumed that the ensemble codes have regular degree distribution of CNs \cite{ashikhmin2004extrinsic}, \cite{ten2004EXIT}.

EXIT function of a CN output (or a VN input in the previous local iteration) and EXIT function of the VN output (or the CN input in the same local iteration),  denoted by $I_{EC}^{(l)}(I_{EV}^{(l)})$ and $I_{EV}^{(l)}(I_{EC}^{(l-1)},\sigma_{ch},p)$ can be found according to our discussion in Section \ref{EXIT-CHART ANALYSIS}.

In order to utilize EXIT chart in designing of LDPC codes for correlated sources, it should be found \emph{two edge-type} coefficients $ \alpha $ and $ \beta $ in addition to $ \lambda(x) $.

By applying a linear programming method, we first design a degree distribution pair $(\lambda, \rho)$ and its corresponding coefficient pair $(\alpha, \beta)$. Thus, the initial value of $(\alpha, \beta)$ is realized and then the $ \{ \lambda_{d_v}, d_v\in (2,\dots,D_v) \} $ is designed again for the check-regular ensemble where the maximum VN degree $D_v$ is predefined. For an ensemble given $\{\lambda_{d_v}\}  $, EXIT function of the VN output can be represented by
\begin{equation*}
I_{EV}=\sum_{d_{v}=2}^{D_v} \lambda_{d_v} I_{EV}(d_v),
\end{equation*}
where $I_{cv}(d_v)$ is EXIT curve of degree $d_v$ VN that is obtained as described in Section \ref{EXIT-CHART ANALYSIS}.

As formerly mentioned, EXIT curve of a CN is the same as that in the P2P case, as follows
\begin{equation*}
I_{E,C}(I_{A,C},d_c)=1 - J\left( \sqrt{(d_c-1)[J^{-1}(1-I_{A,C})]^2} \right),
\end{equation*}
where $ I_{A,C} $ is easily obtained associated to $I_{E,C}$ according to the inverse of $J(.)$ function.

The code rate optimization problem is formulated as follows:
\begin{align*}
\quad \textrm{maximize:}\ \ \ \      &  R=1-\frac{\sum_{d_c=2}^{D_c}\rho_{d_c}/d_c}{\sum_{d_v=2}^{D_{v}}\lambda_{d_v}/d_v} ,\\
\quad \textrm{subject to:} \ \ \ \
      &  1.\ \sum_{d_v=2}^{D_v}\lambda_{d_v} = 1 ,\\
      &  2.\ \lambda_{d_v} \geq 0 ,\\
      &  3.\ I_{EV} > I_{A,C}(I_{E,C}) ,\\
      &\textrm{for}\ I_{E,C} \in [0 \ 1] .
\end{align*}
The third constraint is the zero-error constraint or is equivalent to that the output MI of the VN in each local iteration greater than the previous local iteration (i.e, $I_{EV}^{(l)}>I_{EV}^{(l-1)}$). Also maximizing $R$ is equivalent to maximize $ \sum_{d_v=2}^{D_v}\lambda_{d_v}/d_v $. When $ I_{EV}(d_v) $ is given for each degree so $ I_{EV} $ is linear in $ \{ \lambda_{d_v}, d_v\in (2,\dots,d_{v,max}(D_v)) \} $.

Therefore after finding $ \lambda(x) $ and the given CN degree distribution $ \rho(x) $, we can obtain the optimum $ (\alpha,\beta) $ by using Conditions (\ref{condition-alpha-beta}), (\ref{condition-alpha}) and a stability condition that is given under Gaussian assumption and using MI evolution as \cite{poulliat2005analysis}
\begin{equation}
{\alpha_2 \lambda_2}e^{(\frac{-M^2}{8})} + {(1-\alpha_2) \lambda_2} < \frac{e^{\frac{1}{2\sigma_n^2}}}{\sum_{j=2}^{D_{c}}{\rho_j (j-1)}} ,
\end{equation}
where $ M = J^{-1}( \tilde{J}(\sigma_{max},p) ) $ and $ \sigma_{max}=J^{-1}(1) $. Note that the stability condition depends on the channel parameters and $ \tilde{J}(.) $ function.

The result of our search for the BIAWGN channel is summarized in Table \ref{T-opt-0.9} and Table \ref{T-opt-0.95}.  Table \ref{T-opt-0.9} and Table \ref{T-opt-0.95} contains those degree distribution pairs of rate one-half with coefficient pairs ($\alpha,\beta$) and various correlation parameter. We also consider an upper bound from (\ref{shannon-SW-limit}) which is useful to measure the performance of our designed ensembles.
\begin{table*}[!ht]
\centering
\caption{Good degree distribution pairs ($\lambda,\rho$) of rate one-half with coefficient pairs ($\alpha,\beta$) and correlation parameter $p=0.9$ for the iterative joint decoder over the BI-AWGN channel. Maximum variable node degrees are $d_v=4,5,15,30,45$. For each degree distribution pair the threshold value $(\frac{E_{so}}{N_0})^*$, and the gap between Shannon-SW limit and threshold value are given.}
\label{T-opt-0.9}
\begin{tabular}{|c|c|c||c|c||c|c||c|c||c|c|}
\hline
\scriptsize $d_v$ &\scriptsize $\lambda_i$ &\scriptsize $\alpha_i$	&\scriptsize $\lambda_i$ &\scriptsize $\alpha_i$	&\scriptsize $\lambda_i$ &\scriptsize $\alpha_i$ &\scriptsize $\lambda_i$&\scriptsize $\alpha_i$ &\scriptsize $\lambda_i$    &\scriptsize $\alpha_i$  	\\ \hline

\scriptsize $2$ & \tiny $0.42126$ & \tiny $0.37260$ & \tiny $0.29862$& \tiny $0.19120$ & \tiny $0.23559$ & \tiny $0.29450$& \tiny $0.19560$& \tiny $0.27020$ & \tiny $0.16860$ & \tiny $0.27560$   \\ \hline

\scriptsize $3$&\tiny $0.53604$&\tiny $0.62686$& \tiny $0.32819$&\tiny $0.78327$ &\tiny $0.39783$ &\tiny $0.67584$ &\tiny $0.36002$&\tiny $0.68632$&\tiny $0.31970$ & \tiny $0.67003$ \\ \hline

\scriptsize $4$&\tiny $0.04270$&\tiny  $0.88973$ &  &   &   &   &   &    &   &       \\ \hline

\scriptsize $5$&   &   & \tiny $0.37319$&\tiny  $0.70251$ &  &   &   &   &   &           \\ \hline

\scriptsize $7$&   &   &   &  &\tiny  $0.14198$&\tiny $0.52780$   &   &   &   &          \\ \hline

\scriptsize $8$ &   &    &   &   &   &  &\tiny $0.00525$  &\tiny $0.50105$  &  &    \\ \hline

\scriptsize $9$ &   &    &   &   &   &  &\tiny  $0.21627$  &\tiny  $0.53864$ &\tiny $0.22443$ & \tiny $0.67003$ \\ \hline

\scriptsize $14$   &   &   &   &   &\tiny  $0.00148$ &\tiny $0.50014$    &   &   &  &   \\ \hline

\scriptsize $15$  &   &   &   &   &\tiny  $0.22312$  &\tiny $0.52040$    &   &   &   &   \\ \hline

\scriptsize $29$ &   &    &   &   &   &  &\tiny  $0.00226$  &\tiny $0.50013$   &    &  \\ \hline

\scriptsize $30$ &   &    &   &   &   &  &\tiny $0.22060$  &\tiny $0.38650$   &   &    \\ \hline

\scriptsize $44$ &   &   &   &    &   &   &   &   &\tiny $0.00188$  & \tiny $0.50007$  \\ \hline

\scriptsize $45$  &   &   &   &    &   &   &   &   &\tiny $0.28539$ & \tiny $0.46210$ \\ \hline \hline

\rule{0pt}{0.9\normalbaselineskip} & \multicolumn{2}{c|}{\scriptsize $\rho(x) = x^4$} & \multicolumn{2}{c|}{\scriptsize $\rho(x) = x^5$} & \multicolumn{2}{c|}{\scriptsize $\rho(x) = x^6$} & \multicolumn{2}{c|}{\scriptsize $\rho(x) = x^7$} & \multicolumn{2}{c|}{\scriptsize $\rho(x) = x^8$}            \\ \hline

\scriptsize $k$ & \multicolumn{2}{c|}{\scriptsize $\beta_i$ }  &\multicolumn{2}{c|}{\scriptsize $\beta_i$}                       & \multicolumn{2}{c|}{\scriptsize $\beta_i$ }                    & \multicolumn{2}{c|}{\scriptsize $\beta_i$ }                   & \multicolumn{2}{c|}{\scriptsize $\beta_i$ }              \\ \hline

\scriptsize $1$ & \multicolumn{2}{c|}{\tiny $0.09266$}   &  \multicolumn{2}{c|}{\tiny $0.140400$}  & \multicolumn{2}{c|}{\tiny $0.120379$}  & \multicolumn{2}{c|}{\tiny $0.122219$} & \multicolumn{2}{c|}{\tiny $0.073260$}\\ \hline

\scriptsize $2$  &  \multicolumn{2}{c|}{\tiny $0.26633$}  &\multicolumn{2}{c|}{\tiny $0.197100$}  & \multicolumn{2}{c|}{\tiny $0.126914$}  & \multicolumn{2}{c|}{\tiny $0.120999$} &  \multicolumn{2}{c|}{\tiny $0.085869$}\\ \hline

\scriptsize $3$ &  \multicolumn{2}{c|}{\tiny $0.53459$} & \multicolumn{2}{c|}{\tiny $0.125800$} & \multicolumn{2}{c|}{\tiny $0.133757$} & \multicolumn{2}{c|}{\tiny $0.114674$} & \multicolumn{2}{c|}{\tiny $0.094437$} \\ \hline

\scriptsize $4$  & \multicolumn{2}{c|}{\tiny $0.10650$} & \multicolumn{2}{c|}{\tiny $0.137459$} & \multicolumn{2}{c|}{\tiny $0.201200$} & \multicolumn{2}{c|}{\tiny $0.149333$} &  \multicolumn{2}{c|}{\tiny $0.111274$} \\ \hline

\scriptsize $5$ & \multicolumn{2}{c|}{} &  \multicolumn{2}{c|}{\tiny $0.399241$}  & \multicolumn{2}{c|}{\tiny $0.376500$}  & \multicolumn{2}{c|}{\tiny $0.252000$} &  \multicolumn{2}{c|}{\tiny $0.253140$}  \\ \hline

\scriptsize $6$ & \multicolumn{2}{c|}{} & \multicolumn{2}{c|}{} & \multicolumn{2}{c|}{\tiny $0.041250$}  & \multicolumn{2}{c|}{\tiny $0.207650$}  &  \multicolumn{2}{c|}{\tiny $0.292400$} \\ \hline

\scriptsize $7$ &  \multicolumn{2}{c|}{}  & \multicolumn{2}{c|}{}  & \multicolumn{2}{c|}{}  & \multicolumn{2}{c|}{\tiny$0.033125$} &  \multicolumn{2}{c|}{\tiny $0.076500$}  \\ \hline

\scriptsize $8$  &  \multicolumn{2}{c|}{}   & \multicolumn{2}{c|}{}  & \multicolumn{2}{c|}{}  & \multicolumn{2}{c|}{} &  \multicolumn{2}{c|}{\tiny$0.013120$} \\ \hline \hline

\tiny $(\frac{E_{so}}{N_0})_{dB}^*$  & \multicolumn{2}{c|}{\scriptsize $-0.53$}  &\multicolumn{2}{c|}{\scriptsize $-0.94$}  & \multicolumn{2}{c|}{\scriptsize $-1.32$}  & \multicolumn{2}{c|}{\scriptsize $-1.45$} & \multicolumn{2}{c|}{\scriptsize $-1.52$} \\ \hline \hline

\scriptsize  gap(dB)  &  \multicolumn{2}{c|}{\scriptsize $1.24$}  & \multicolumn{2}{c|}{\scriptsize $0.83$}   & \multicolumn{2}{c|}{\scriptsize $0.45$}   & \multicolumn{2}{c|}{\scriptsize $0.32$}  & \multicolumn{2}{c|}{\scriptsize $0.25$}\\ \hline
\end{tabular}
\end{table*}

\begin{table*}[!ht]
\centering
\caption{Good degree distribution pairs ($\lambda,\rho$) of rate one-half with coefficient pairs ($\alpha,\beta$) and correlation parameter $p=0.95$ for the iterative joint decoder over the BI-AWGN channel. Maximum variable node degrees are $d_v=4,5,15,30,48$. For each degree distribution pair the threshold value $(\frac{E_{so}}{N_0})^*$, and the gap between Shannon-SW limit and threshold value are given.}
\label{T-opt-0.95}
\begin{tabular}{|c|c|c||c|c||c|c||c|c||c|c|}
\hline
\scriptsize $d_v$ &\scriptsize $\lambda_i$ &\scriptsize $\alpha_i$	&\scriptsize $\lambda_i$ &\scriptsize $\alpha_i$	&\scriptsize $\lambda_i$ &\scriptsize $\alpha_i$ &\scriptsize $\lambda_i$&\scriptsize $\alpha_i$ &\scriptsize $\lambda_i$    &\scriptsize $\alpha_i$  	\\ \hline

\scriptsize $2$  & \tiny $0.43235$   & \tiny $0.35374$  & \tiny $0.30450$ & \tiny $0.19025$  &  \tiny $0.25180$  & \tiny $0.20240$ & \tiny $0.21190$ & \tiny $0.19450$ & \tiny $0.17742$  & \tiny $0.19526$   \\ \hline

\scriptsize $3$   & \tiny $0.50260$ &\tiny $0.65136$  & \tiny $0.31478$  &\tiny $0.78999$ & \tiny $0.38081$   &\tiny $0.79432$  &\tiny $0.34978$  &\tiny $0.76806$  &\tiny $0.32998$ &\tiny $0.74604$ \\ \hline

\scriptsize $4$  &\tiny $0.06505$ &\tiny  $0.88408$    &   &   &   &   &    &   &   & \\ \hline

\scriptsize $5$  &   &  & \tiny $0.38072$ &\tiny  $0.71962$    &   &   &   &   &    &  \\ \hline

\scriptsize $7$  &   &   &   &   &\tiny  $0.10944$  &\tiny $0.54110$    &   &   &  &  \\ \hline

\scriptsize $8$  &   &    &   &   &   &    &\tiny $0.01869$  &\tiny $0.50589$    &   &   \\ \hline

\scriptsize $9$  &   &    &   &   &   &   &\tiny  $0.14219$  &\tiny  $0.53974$  &\tiny $0.14339$ & \tiny $0.46256$   \\ \hline

\scriptsize $10$   &   &   &   &   &   &   &   &  &\tiny $0.00153$ & \tiny $0.50037$ \\ \hline

\scriptsize $14$  &   &   &   &   &\tiny  $0.00121$   &\tiny $0.50023$    &   &   &  &    \\ \hline

\scriptsize $15$  &   &   &   &   &\tiny  $0.25674$   &\tiny $0.46750$    &   &   &  &   \\ \hline

\scriptsize $29$ &   &    &   &   &   &   &\tiny  $0.00235$ &\tiny $0.50020$  &   &   \\ \hline

\scriptsize $30$   &   &    &   &   &   &   &\tiny $0.27509$  &\tiny $0.54750$  &  &  \\ \hline

\scriptsize $47$  &   &   &   &    &   &   &   &   &\tiny $0.01012$ & \tiny $0.50052$    \\ \hline

\scriptsize $48$   &   &   &   &    &   &   &   &   &\tiny $0.33756$ & \tiny $0.56750$   \\ \hline \hline

\rule{0pt}{0.9\normalbaselineskip}  & \multicolumn{2}{c|}{\scriptsize $\rho(x) = x^4$} & \multicolumn{2}{c|}{\scriptsize $\rho(x) = x^5$} & \multicolumn{2}{c|}{\scriptsize $\rho(x) = x^6$} & \multicolumn{2}{c|}{\scriptsize $\rho(x) = x^7$} & \multicolumn{2}{c|}{\scriptsize $\rho(x) = x^8$}            \\ \hline

\scriptsize $k$ & \multicolumn{2}{c|}{\scriptsize $\beta_i$ }  &\multicolumn{2}{c|}{\scriptsize $\beta_i$}                       & \multicolumn{2}{c|}{\scriptsize $\beta_i$ }                    & \multicolumn{2}{|c|}{\scriptsize $\beta_i$ }                   & \multicolumn{2}{c|}{\scriptsize $\beta_i$ }              \\ \hline

\scriptsize $1$ & \multicolumn{2}{c|}{\tiny $0.105331$} & \multicolumn{2}{c|}{\tiny $0.158400$} &  \multicolumn{2}{c|}{\tiny $0.114195$}  & \multicolumn{2}{c|}{\tiny $0.053840$} &  \multicolumn{2}{c|}{\tiny $0.052944$}   \\ \hline

\scriptsize $2$  & \multicolumn{2}{c|}{\tiny $0.256736$} & \multicolumn{2}{c|}{\tiny $0.004700$} &  \multicolumn{2}{c|}{\tiny $0.116914$}  & \multicolumn{2}{c|}{\tiny $0.066000$} &  \multicolumn{2}{c|}{\tiny $0.064394$} \\ \hline

\scriptsize $3$  &  \multicolumn{2}{c|}{\tiny $0.481433$} & \multicolumn{2}{c|}{\tiny $0.294200$} & \multicolumn{2}{c|}{\tiny $0.119691$} & \multicolumn{2}{c|}{\tiny $0.083500$} & \multicolumn{2}{c|}{\tiny $0.073318$} \\ \hline

\scriptsize $4$  & \multicolumn{2}{c|}{\tiny $0.156500$} & \multicolumn{2}{c|}{\tiny $0.280521$} &  \multicolumn{2}{c|}{\tiny $0.302500$}  & \multicolumn{2}{c|}{\tiny $0.344100$} &  \multicolumn{2}{c|}{\tiny $0.092279$}    \\ \hline

\scriptsize $5$  &  \multicolumn{2}{c|}{} &  \multicolumn{2}{c|}{\tiny $0.262179$}  & \multicolumn{2}{c|}{\tiny $0.264200$}  & \multicolumn{2}{c|}{\tiny $0.177750$} & \multicolumn{2}{c|}{\tiny $0.315045$} \\ \hline

 \scriptsize $6$ &  \multicolumn{2}{c|}{}  & \multicolumn{2}{c|}{} & \multicolumn{2}{c|}{\tiny $0.082500$}  & \multicolumn{2}{c|}{\tiny $0.242309$}   & \multicolumn{2}{c|}{\tiny $0.292400$}\\ \hline

 \scriptsize $7$ &  \multicolumn{2}{c|}{} & \multicolumn{2}{c|}{}  & \multicolumn{2}{c|}{} & \multicolumn{2}{c|}{\tiny$0.032500$} &   \multicolumn{2}{c|}{\tiny $0.076500$}   \\ \hline

\scriptsize $8$ & \multicolumn{2}{c|}{} & \multicolumn{2}{c|}{}  & \multicolumn{2}{c|}{}  &  \multicolumn{2}{c|}{}                 & \multicolumn{2}{c|}{\tiny$0.033120$} \\ \hline \hline

\tiny $(\frac{E_{so}}{N_0})_{dB}^*$ & \multicolumn{2}{c|}{\scriptsize $-1.41$}& \multicolumn{2}{c|}{\scriptsize $-1.64$} & \multicolumn{2}{c|}{\scriptsize $-1.99$} & \multicolumn{2}{c|}{\scriptsize $-2.21$}  &  \multicolumn{2}{c|}{\scriptsize $-2.3$} \\ \hline \hline

\scriptsize gap(dB) & \multicolumn{2}{c|}{\scriptsize $1.08$} & \multicolumn{2}{c|}{\scriptsize $0.85$} & \multicolumn{2}{c|}{\scriptsize $0.5$}  & \multicolumn{2}{c|}{\scriptsize $0.28$} & \multicolumn{2}{c|}{\scriptsize $0.19$} \\ \hline
\end{tabular}
\end{table*}

\section{FINITE-LENGTH RESULTS}\label{finite result}
In this section, we illustrate the performance of finite-length LDPC codes constructed from optimized irregular degree distributions for iterative joint channel decoder at different rates. The finite-length construction is performed with a modified progressive edge growth (PEG) method that has very low error-floor performance \cite{khazraie2012peg}.

We simulate the proposed scheme for different values of the correlation parameters $ p $ and assume that both of transmit nodes use the same degree distribution of LDPC codes and also each of the channel parameters are the same. For the given $p$ and $R$, the theoretical bound of the $E_{so}/N_0$ can be calculated from (\ref{shannon-SW-limit}) for error-floor recovery.

In the following, we represent sample simulation results associated with several designed LDPC codes  for different values of the correlation parameters and the code rates in section \ref{optimization}. We show the performance of the designed codes in section \ref{optimization} are bater than the obtained results in \cite{daneshgaran2006ldpc} by using of the simulation results with finite-length. We consider the ensemble codes with rate $ R=0.5 $ for several the correlation parameter $p$. The blocks length are selected to be $ 20000 $ and the maximum number of local and global iteration are set to 100 and 10 respectively. We provide our simulation results by using the irregular LDPC codes according to Table \ref{T-opt-0.9} and \ref{T-opt-0.95}.\\ \\
\textit{Example} 1: From Table \ref{T-opt-0.9}, we consider the designed code for $R=0.5$ and $p=0.9$ with following VN and CN degree distribution and its corresponding the pair $ (\alpha,\beta) $
\begin{align*}
\begin{split}
\lambda(x)= & 0.23559 \ x + 0.39783 \ x^3 + 0.14198 \ x^6 + \\
& 0.00148 \ x^{13} + 0.22312 \ x^{14},   \\
\end{split}
\end{align*}
\begin{align*}
\rho(x)= x^6.
\end{align*}
\\ \\
\textit{Example} 1: From Table \ref{T-opt-0.95}, we consider the designed code for $R=0.5$ and $p=0.95$ with following VN and CN degree distribution and its corresponding the pair $ (\alpha,\beta) $
\begin{align*}
\begin{split}
\lambda(x)= & 0.2518 \ x + 0.38081 \ x^3 + 0.10944 \ x^6 + \\
& 0.00121 \ x^{13} + 0.25674 \ x^{14},   \\
\end{split}
\end{align*}
\begin{align*}
\rho(x)= x^6.
\end{align*}

For the correlated sources, the BER values of three codes as a function of $ E_{so}/N_0 $ (SNR) and $ p $ are reported in Fig. \ref{F-simulation-0.5}. Note that the curves labeled ``glob.it.0'' show the performance of the LDPC without using the correlation information that are equal to the performance of point to point.

\begin{figure}[!t]
\centering
\includegraphics[scale=0.65]{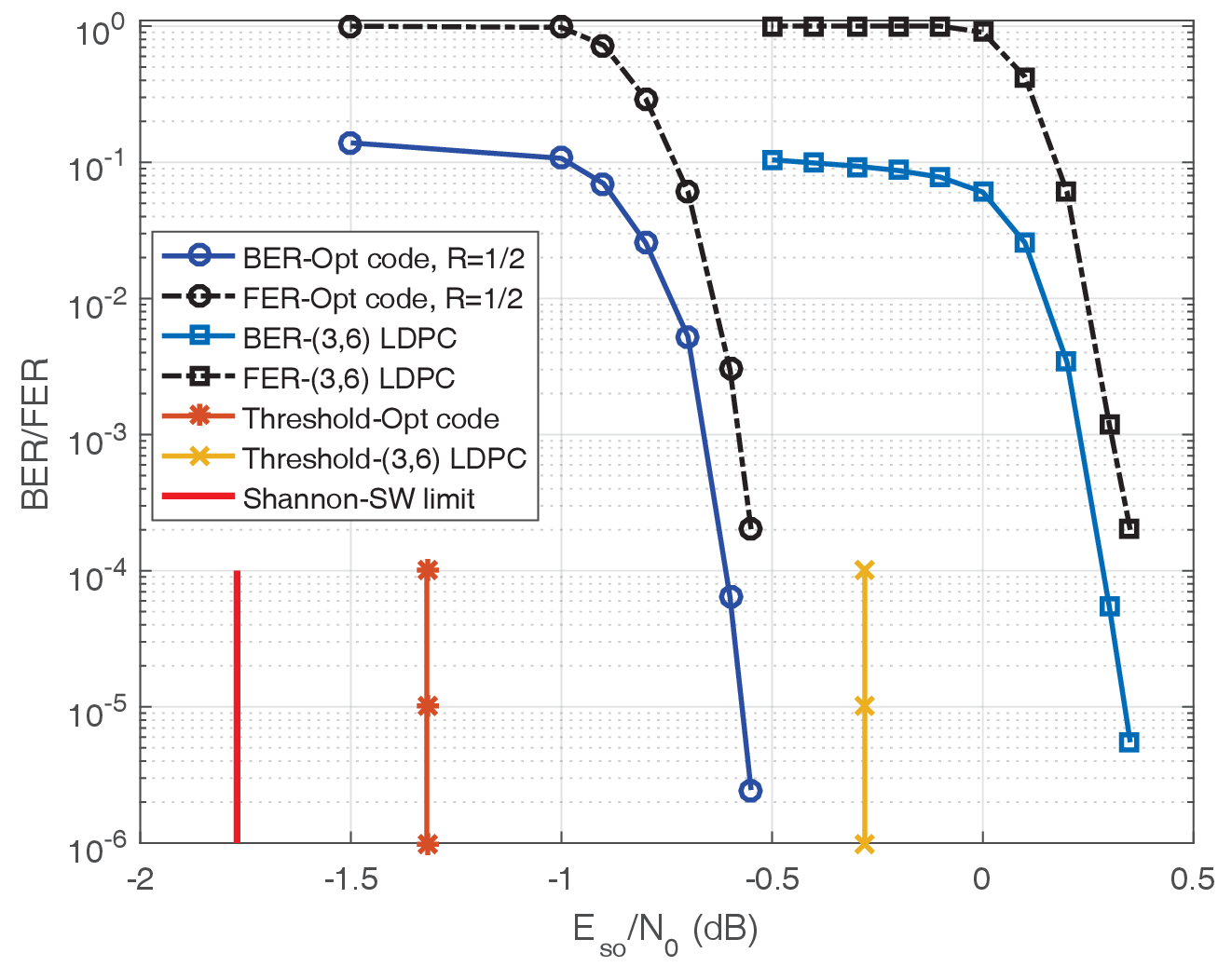}
\caption{correlation $0.9$}\label{F-EXIT36}
\end{figure}

\begin{figure}[!t]
\centering
\includegraphics[scale=0.65]{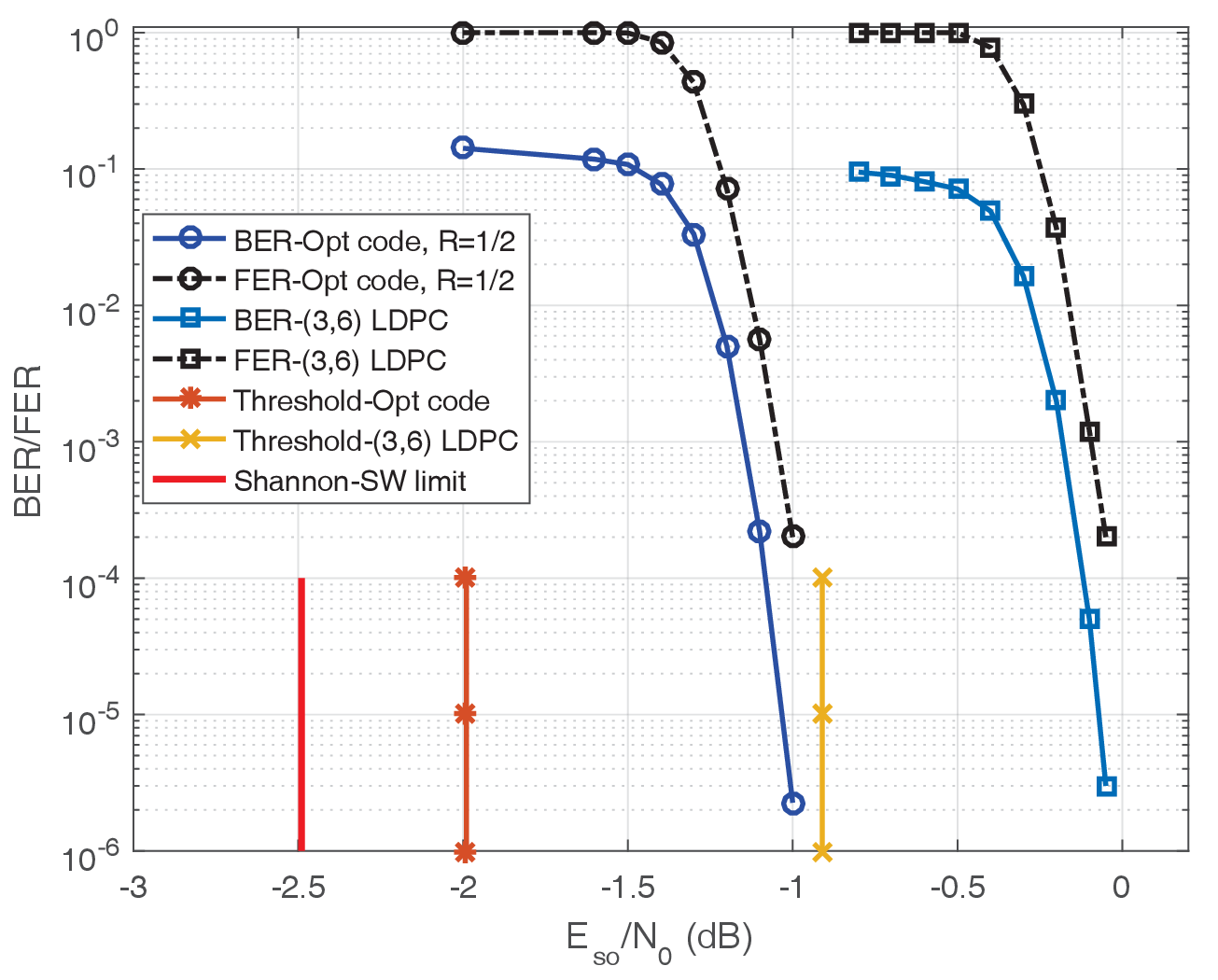}
\caption{ correlation $0.95$}\label{F-EXIT36}
\end{figure}

Table \ref{table-3} shows, for various values of the correlation parameter $p$ with the constant rate $R=0.5$, the corresponding joint entropy $ H(u_1,u_2) $ of the correlated sources, the theoretical limit for $ E_{so}/N_0 $ in (\ref{shannon-SW-limit}), $ [E_{so}/N_0]_{lim} $, the obtained threshold from EXIT chart, $ [E_{so}/N_0]_{th} $, and the value of $ E_{so}/N_0 $ for which the proposed iterative joint decoder (at target
BER$=10^{-5}$).

\section{CONCLUSIONS}\label{Conclusion}


\bibliographystyle{IEEEtran}
\bibliography{IEEEabrv,biblio2}

\end{document}